\documentclass[preprint2]{aastex}

\usepackage{longtable}
\usepackage{amsmath}

\shorttitle{Photoionization and photodissociation of toluene}
\shortauthors{Monfredini et al.}

\begin{document}


\title{Single and Double Photoionization and Photodissociation of Toluene by Soft X-rays in Circumstellar Environment}



\author{T. Monfredini$^{1}$, F. Fantuzzi$^{2}$, M. A. C. Nascimento$^{2}$, W. Wolff$^{3}$, and H. M. Boechat-Roberty$^{1,*}$}

\affil{$^{1}$Observat\'{o}rio do Valongo, Universidade Federal do Rio de Janeiro, Ladeira Pedro Antonio, 43, Rio de Janeiro, Brazil}

\affil{$^{2}$Instituto de Qu\'{i}mica, Universidade Federal do Rio de Janeiro, Av. Athos da Silveira Ramos, 149, Rio de Janeiro, Brazil}

\affil{$^{3}$Instituto de F\'{i}sica, Universidade Federal do Rio de Janeiro, Av. Athos da Silveira Ramos, 149, Rio de Janeiro, Brazil}

\email{*Email: heloisa@astro.ufrj.br; Tel: +55 21 2263 0685 r.216}




\begin{abstract}
The formation of polycyclic aromatic hydrocarbons (PAHs) and their methyl derivatives occurs mainly in the dust shells of asymptotic giant branch (AGB) stars. The bands at 3.3 and 3.4 $\mu$m, observed in infrared emission spectra of several objects, are attributed C-H vibrational modes in aromatic and aliphatic structures, respectively. In general, the feature at 3.3 $\mu$m is more intense than the 3.4 $\mu$m. Photoionization and photodissociation processes of toluene, the precursor of methylated PAHs, were studied using synchrotron radiation at soft X-ray energies around the carbon K edge with time-of-flight mass spectrometry. Partial ion yields of a large number of ionic fragments were extracted from single and 2D-spectra, where electron-ion coincidences have revealed the doubly charged parent-molecule and several doubly charged fragments containing seven carbon atoms with considerable abundance. \textit{Ab initio} calculations based on density functional theory were performed to elucidate the chemical structure of these stable dicationic species. The survival of the dications subjected to hard inner shell ionization suggests that they could be observed in the interstellar medium, especially in regions where PAHs are detected. The ionization and destruction of toluene induced by X-rays were examined in the T Dra conditions, a carbon-rich AGB star. In this context, a minimum photodissociation radius and the half-life of toluene subjected to the incidence of the soft X-ray flux emitted from a companion white dwarf star were determined.
\end{abstract}





\section{Introduction}

Toluene (C$_6$H$_5$CH$_3$), or methyl-benzene, is the simplest alkyl-substituted benzene derivative, in which a methyl group replaces one of the hydrogen atoms of the benzene (C$_6$H$_6$) molecule. Although these species have not yet been detected in the interstellar medium, its protonated version (C$_7$H$_9^+$) has a spectral signature that matches reasonably with the unidentified interstellar infrared emission band near 6.2 $\mu$m \citep{dou08}. Moreover, the mechanisms of formation and the role of the molecule in astrophysical environments and in the combustion chemistry have attracted considerable attention in the last years (Dagaut et al. 2002; Imanaka \& Smith 2007; Silva et al. 2007; Dangi et al. 2013).

Toluene was firstly identified as one of the major constituents of the Murchison and Allende meteorites \citep{stu72}. A catalytic combination of CO, H$_2$ and NH$_3$ on the surface of dust grains - a process known as Fischer-Tropsch-type reaction - is attributed as the main formation route of toluene and the meteorite organic matter \citep{stu72}. Toluene is also considered a critical building block for the formation of methyl-substituted polycyclic aromatic hydrocarbons (PAHs) (Thomas \& Wornat 2008; Ricks et al. 2009), which has been proposed as the carriers of the 3.4 $\mu$m emission in astrophysical environments. It is well known that the formation of PAHs and their methyl derivatives occurs mainly in the dust shells of asymptotic giant branch (AGB) stars, which after the ejection of its carbon-rich envelope into the interstellar medium become protoplanetary nebulae \citep{che92}. In fact, the infrared emission features that prevail over the spectra of most galactic and extragalactic sources are carried by these molecules \citep{tie08}.

Concerning the astrophysical synthesis of methylated benzene, different routes of formation are proposed in the literature, but yet no consensus was reached. Tokmakov et al. (1999), combining experimental measurements and theoretical methods, investigated the reaction of the phenyl radical (C$_6$H$_5$) with methane, and concluded that this route does not lead preferentially to toluene, but mainly to benzene and CH$_3$. More recently, \citep{dan13} showed that the gas phase synthesis of toluene from the ethynyl radical (C$_2$H) with isoprene (2-methyl-1,3-butadiene, C$_5$H$_8$) - two neutral, non-cyclic precursors - has no entrance barrier and all the transition states have energy lower than the separated reactants. Consequently, it is a viable reaction pathway in low-temperature environments, such as the cold molecular cloud TMC-1. Another possible route of formation of interstellar toluene is the reaction of o-benzyne (C$_6$H$_4$) with methane, but to the best of our knowledge this reaction was neither experimentally nor theoretically studied.

Ziegler et al. (2005) proposed that the formation process of toluene in the gas-phase could take place in the warm layers of the circumstellar envelope of AGB stars, involving the reaction of benzyl with H and l-C$_{3}$H$_{3}$ with C$_{4}$H$_{5}$. The object T Dra is a carbon-rich AGB star associated with FUV and X-ray emission. The origin of the X-ray emission in the environment of the star, however, is still a matter of debate. It was proposed that the presence of a white dwarf as a companion, in a binary system, could be responsible for the observable X-ray flux \cite{Sahai2008, Ramstedt2012}. However, more observations are necessary to constrain the intensity and periodicity of these events in T Dra and other AGB stars, as well as the role in its inner shell. Nevertheless, the presence of a companion star should not be considered surprising, since it has been proposed for well-studied evolved stars, such as IRC +10216 and the pre-planetary nebula CRL 618 \cite{Velazquez2014, Cernicharo2015, Kim2015}.

The present work obtained experimental and theoretical information about the photoionization and photodissociation processes of toluene upon interaction with soft X-ray photons at energies in the vicinity of the carbon K shell. Particular attention was given to the analysis of doubly charged species, inasmuch as the presence of multiply-charged PAHs in the interstellar medium has been proposed \cite{Rosi2004,Malloci2007}. Since the chemistry of circumstellar regions is strongly affected by X-rays photons associated with  high velocity winds and with shocked gas in a binary companion, the investigation of the photo-interaction is of great interest for unravelling the behavior of molecules in these environments. Here, the emphasis is placed on the environment of the AGB stellar object T Dra, and the spotlight of the work is directed to determine the ionization, destruction, and the half-life of toluene under the observed astrophysical conditions of the star. We expect that our results will open a discussion towards the presence of ion-molecule reactions and the structure of molecules in different charged states in the circumstellar envelope of AGB stars.

In the following section, we briefly present the experimental and theoretical methods used in this work. Section 3 contains the results and discussion. In the last section, final remarks and conclusions are presented.

\section{Methodology}

The measurements were performed at the Brazilian Synchrotron Light Source Laboratory (LNLS) using soft X-rays photons selected from the Spherical Grating Mono-chromator (SGM) at energies around the C1s resonance (280-320 eV). The experimental set-up has been described in detail elsewhere \cite{boe05}. Briefly, soft X-ray photons, at a rate of ~10$^{12}$ photons s$^{-1}$, perpendicularly intersect the gas sample inside a high vacuum chamber. The photons absorbed by the molecules produce the excitation and ionization of electrons from inner shell, causing a molecular instability that can lead to the dissociation, producing several ionized fragments. The base pressure in the vacuum chamber was maintained in the 10$^{-8}$ Torr range, while during operation the chamber pressure was raised up to 10$^{-6}$ Torr.  The gas needle was kept at ground potential. The emerging photon beam was recorded by a light sensitive diode. The sample was purchased commercially from Sigma-Aldrich with purity better than 99.5\%. No further purification was used except for degassing the liquid sample by multiple freeze-pump-thaw cycles before admitting the vapor into the chamber. Since toluene possesses a high vapor pressure, there was no need to heat the liquid sample, avoiding degradation.

The photoelectrons (PE) and photoions (PI) produced by the interaction with the photon beam were extracted by a static electric field of 450 V/cm. The ionized fragments were accelerated and focused by a two-stage electric field into a field free drift tube 297 mm long and detected by micro-channel plate detector in a chevron configuration. The ionic species were mass-to-charge (m/z) analyzed by a time-of-flight mass spectrometer (TOF-MS), and produced the stop signals delivered to a time-to-digital converter (TDC). The photoelectrons were directed into the opposite direction towards a micro-channel plate detector. The TDC was started by one of the collected electrons. Standard time-of-flight spectra were obtained using the correlation between one photoelectron and a photoion in coincidence (PEPICO). In addition to the PEPICO spectra, two-dimensional coincidence mass spectra (PE2PICO: Photo-Electron Photo-ion Photo-ion Coincidence) were simultaneously obtained. In this case, the ion pairs resulted from the multiple ionization events, mostly associated with the Auger-like process, that after decay break into at least two ionic and neutral species. Only about 0.6\% of all coincidence signals were sorted in the PE2PICO spectra, reflecting that the majority of the contribution was correlated to single coincidence events.

In order to extract astrochemical informations it is essential to obtain the absolute ionization and fragmentation cross sections. The data reduction procedure included subtraction of a linear background and the false coincidences coming from aborted double and triple ionization events \cite{simon1991}. Then, contributions of all cationic fragments were summed up and normalized to the photoabsorption cross sections measured by Ishii \& Hitchcock (1987). Assuming a negligible fluorescence yield due to the low carbon atomic number \cite{chen1981} and an ionic fragment production in the present photon energy range, it is possible to consider that all absorbed photons lead to cation formation. Inner shell photoabsorption and the subsequent photoionization process may produce instabilities on the molecular structure (nuclear rearrangements), producing a transient molecule that breaks down following different dissociation pathways. Mass-spectra of toluene were obtained around the energy of the resonance C1s $\rightarrow \pi^{*}$ \cite{hit94} between 277.9 and 302.7 eV.

Density functional calculations were performed to elucidate the most stable molecular structures of the doubly ionized species obtained in the X-rays photodissociation measurements. These ions are of type C$_7$H$_n^{++}$, n=2-8. Around 70 initial geometries for each dication were optimized at the M06-2X/cc-pVTZ(-f) level (Zhao \& Truhlar 2008). Frequency calculations and thermodynamic analysis were performed for the 15 most stable structures of each species, and more than 100 different minimum structures were characterized. Considering this large number, the structural and thermodynamic aspects were discussed solely on the basis of the most stable C$_7$H$_n^{++}$ species, at the mentioned level of theory. The low-lying isomers of each C$_7$H$_n^{++}$ dication will be discussed in a future work.  Moreover, the thermodynamics of several dissociation pathways of the dications by Coulombic decay were evaluated, and compared to the estimated coincidence of the monocation products from the PE2PICO analysis. All calculations were performed with the Jaguar 7.9 program.      

\begin{figure*}[t!]
\centering
  \includegraphics[width=130mm]{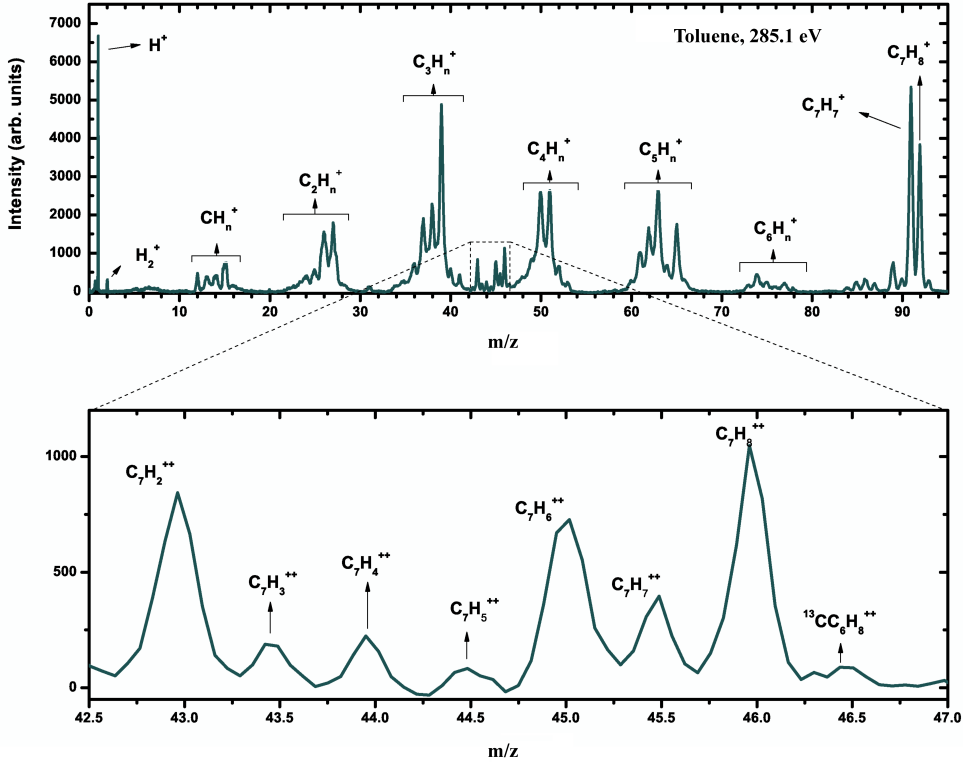}
\caption{Time-of-flight mass spectrum of toluene after exposition to X-ray at 285.1 eV.} \label{spec285}
\end{figure*}

The radiative flux model of T Dra was constructed by using the numerical simulations of wind accretion in symbiotic binary systems developed by Val-Borro et al. (2009). The authors estimated an X-ray luminosity due to mass accretion of the wind of an AGB star onto a small companion (R=0.02 R$_{\odot}$), between distances of 16 to 70 AU, in the range between 3$\times$10$^{30}$ and 3$\times$10$^{32}$ erg s$^{-1}$. This scenario was applied to obtain the photoionization and photodissociation rates of toluene, as well as the photoproduction of dications, as functions of the distance from the AGB star.

\section{Results and Discussion}

\subsection{Experimental results}

Figure \ref{spec285} shows the mass spectra of toluene at the C1s resonance energy (285.1 eV). Besides the C$_{7}$ group (84 to 93 amu), the following ionic groups were identified, due to carbon atom loss: methyl (12 to 16 amu), C$_{2}$ (24 to 28 amu), C$_{3}$ (36 to 41 amu), C$_{4}$ (48 to 53 amu), C$_{5}$ (60 to 66 amu), and C$_{6}$ (73 to 78 amu). Moreover, the appearance of fractional m/z from 42 to 47 amu suggests that doubly ionized molecules from the C$_{7}$ group were also produced. The region of the spectra corresponding to the double ionized molecules from the C7 group, of special interest in this work, is enhanced in Figure \ref{spec285}, and the fragment-ions assignments highlighted. It should be mentioned that such a full sequence of doubly ionized molecules is not usually encountered and easily identified in homo and heterocyclic ring molecules. Small features at masses (m/z) 20, 38.5 and 39.5, corresponding to the ions C$_{3}$H$_{4}^{++}$, C$_{6}$H$_{3}^{++}$ and C$_{6}$H$_{5}^{++}$, were also identified. Small doubly charged molecules, in general, are metastable: charge separation may occur and the fragments are accelerated by their mutual repulsive character (Coulomb explosion). It is well established, however, that aromatic molecules resist Coulomb explosion better than other molecules \cite{Burdick1986}.

Even though the discrimination of doubly charged ions in mass spectra can be difficult or even impossible, they are very easily identified in the mass spectra of toluene. Shaw et al (1998) reported the appearance of the ion C$_{7}$H$_{8}^{++}$ with ionization by photons of energy 23.8 eV. The peaks of the double ionized fragments in our mass-spectra present a narrower profile than those for small single ionized fragments, which reflects the thermal energy of the ions. Doubly ionized peaks of C$_{7}$H$_{n}^{++}$, with n = 2-8, were also produced from the fragmentation of toluene with the impact of 100 eV electrons (G\l{}uch et al. 2005)  

\begin{table}[t!]
\begin{footnotesize}
\centering
\caption{Partial Ion Yield (PIY) for the fragmentation of toluene in the gas phase, due to photons with energies around the C1s resonance.}
\label{tab:PIY}
\begin{tabular}{lllllll}
\hline
\hline
\multicolumn{2}{c}{Fragments} & \multicolumn{5}{c}{PIY (\%) per energy (eV)} \\
m/z  & Attr.               & 277.9 & 285.1 & 286.6 & 289.1 & 302.7 \\
\hline
1    & H$^{+}$             & 6.8   & 7.9   & 8.8   & 10.2  & 11.6  \\
13   & CH$^{+}$            & 1.5   & 1.5   & 1.5   & 1.6   & 1.8   \\
14   & CH$^{+}_{2}$        & 1.3   & 1.6   & 1.6   & 1.8   & 2.5   \\
15   & CH$^{+}_{3}$        & 2.4   & 2.2   & 2.1   & 2.3   & 3.7   \\
24   & C$_{2}^{+}$         & 1.4   & 1.9   & 1.7   & 2.0   & 2.3   \\
25   & C$_{2}$H$^{+}$      & 1.1   & 1.4   & 1.5   & 1.6   & 2.4   \\
26   & C$_{2}$H$_{2}^{+}$  & 4.9   & 5.3   & 5.5   & 5.9   & 7.2   \\
27   & C$_{2}$H$_{3}^{+}$  & 4.8   & 4.7   & 4.9   & 4.5   & 7.0   \\
36   & C$_{3}^{+}$         & 1.2   & 2.1   & 1.8   & 2.8   & 2.5   \\
37   & C$_{3}$H$^{+}$      & 3.8   & 4.4   & 5.5   & 5.0   & 5.9   \\
38   & C$_{3}$H$_{2}^{+}$  & 3.5   & 4.1   & 4.2   & 5.2   & 4.8   \\
39   & C$_{3}$H$_{3}^{+}$  & 8.0   & 8.0   & 7.9   & 7.2   & 7.4   \\
43   & C$_{7}$H$_{2}^{++}$ & 0.4   & 0.5   & 0.5   & 0.5   & 0.7   \\
43.5 & C$_{7}$H$_{3}^{++}$ & 0.05  & 0.1   & 0.1   & 0.1   & 0.15  \\
44   & C$_{7}$H$_{4}^{++}$ & 0.05  & 0.1   & 0.1   & 0.1   & 0.1   \\
44.5 & C$_{7}$H$_{5}^{++}$ & 0.04  & 0.04  & 0.06  & 0.04  & 0.05  \\
45   & C$_{7}$H$_{6}^{++}$ & 0.4   & 0.4   & 0.4   & 0.4   & 0.5   \\
45.5 & C$_{7}$H$_{7}^{++}$ & 0.2   & 0.2   & 0.2   & 0.2   & 0.1   \\
46   & C$_{7}$H$_{8}^{++}$ & 0.9   & 0.5   & 0.6  & 0.5   & 0.7   \\
48   & C$_{4}^{+}$         & 0.8   & 0.9   & 1.1   & 1.6   & 1.6   \\
49   & C$_{4}$H$^{+}$      & 1.7   & 2.6   & 2.5   & 3.1   & 3.2   \\
50   & C$_{4}$H$_{2}^{+}$  & 5.2   & 5.8   & 5.9   & 5.5   & 5.2   \\
51   & C$_{4}$H$_{3}^{+}$  & 4.3   & 4.7   & 4.6   & 5.1   & 5.5   \\
52   & C$_{4}$H$_{4}^{+}$  & 1.6   & 1.6   & 1.4   & 0.7   & 0.6   \\
61   & C$_{5}$H$^{+}$      & 1.2   & 1.9   & 2.2   & 3.1   & 2.0     \\
62   & C$_{5}$H$_{2}^{+}$  & 2.4   & 3.6   & 3.6   & 3.7   & 2.7   \\
63   & C$_{5}$H$_{3}^{+}$  & 3.9   & 4.5   & 4.6   & 4.9   & 4.2   \\
64   & C$_{5}$H$_{4}^{+}$  & 1.8   & 1.7   & 1.7   & 1.0  & 0.41  \\
65   & C$_{5}$H$_{5}^{+}$  & 3.2   & 2.7   & 2.3   & 1.8   & 1.2   \\
91   & C$_{7}$H$_{7}^{+}$  & 10.6  & 5.8   & 5.3   & 3.3   & 1.4   \\
92   & C$_{7}$H$_{8}^{+}$  & 7.8   & 3.8   & 3.6   & 2.5   & 1.1  \\
\hline
\hline 
\end{tabular}
\end{footnotesize}
\end{table}

As shown by Bakes et al. (2001) in a model under interstellar conditions, the infrared emission from PAH molecules depend on their charge state and temperature distribution function. The production of doubly charged ions from PAHs with more than 24 carbon atoms may be dominant for a G$_{0}$/n$_{e}$ ratio larger than 10$^{3}$ cm$^{-3}$. It also changes the ionization degree of the cloud. In environments with some contribution of soft X-rays, these dications can be produced by the absorption of photons of energy around the C1s edge, followed by Auger process of both neutral (emitting two electrons) and cationic molecules (emitting just one electron). Reitsma et al. (2015) studied the production and stability of dications of coronene from X-ray photon interaction with cations. This may be an important route to produce dications, but it still lacks the rate production of doubly charged ions from neutral PAHs, since part of them is destroyed by photoionization. Rosi et al. (2004) investigated theoretically the stability, structure and fragmentation routes of C$_{6}$H$_{6}^{++}$. Toluene is considered a small molecule, which shows experimental and theoretical advantages to investigate dication stability and production from the neutral parent.

\begin{table*}[t!]
\centering 
\caption{Values of non-dissociative single ($\sigma_{ph-i}$) and double ($\sigma_{ph-ii}$) photoionization cross sections; dissociative ionization (photodissociation) cross section ($\sigma_{ph-d}$); the and cross section of dications production ($\sigma_{++}$) of toluene around C1s edge. The photoabsorption cross section ($\sigma_{ph-abs}$) from Ishii $\&$ Hitchcock (1987) is also shown.} \label{tab-sigma}
\begin{tabular}{l l c c c c}
\\
\hline \hline
Energy (eV)&  \multicolumn{4}{c}{Cross Sections (cm$^{2}$)}\\
	&	 $\sigma_{ph-abs}$	&	 $\sigma_{ph-i}$   	&	 $\sigma_{ph-d}$ 	&	 $\sigma_{ph-ii}$	&	 $\sigma_{++}$	\\
\multicolumn{6}{l}{\textit{Toluene}}\\
277.9	&	 6.2$\times 10^{-20}$	&	  4.8$\times 10^{-21}$  	&	 5.7$\times 10^{-20}$ 	&	  5.7$\times 10^{-22}$  	&	  1.3$\times 10^{-21}$  	\\
285.1	&	 1.9$\times 10^{-17}$	&	  7.3$\times 10^{-19}$  	&	 1.8$\times 10^{-17}$ 	&	  9.6$\times 10^{-20}$  	&	  3.6$\times 10^{-19}$  	\\
286.6	&	 1.8$\times 10^{-18}$	&	  6.4$\times 10^{-20}$  	&	 1.7$\times 10^{-18}$ 	&	  9.9$\times 10^{-21}$  	&	  3.5$\times 10^{-20}$  	\\
289.1	&	 7.3$\times 10^{-18}$	&	  1.8$\times 10^{-19}$  	&	 7.1$\times 10^{-18}$ 	&	  3.4$\times 10^{-20}$  	&	  1.3$\times 10^{-19}$  	\\
302.7	&	 9.5$\times 10^{-18}$	&	  1.0$\times 10^{-19}$  	&	 9.4$\times 10^{-18}$ 	&	  7.0$\times 10^{-20}$  	&	  2.4$\times 10^{-19}$  	\\
\multicolumn{6}{l}{\textit{Benzene} (Boechat-Roberty et al. 2009)}\\
285.1	&	 4.2$\times 10^{-18}$	&	  1.7$\times 10^{-19}$  	&	 3.7$\times 10^{-18}$ 	&	  -  	&	  -  	\\
\hline \hline
\end{tabular}
\end{table*}

As some peak contribution of fragment-ions are superposed, multiple gaussian function profiles were fitted to the peaks in order to extract the partial ion yield (PIY) of each fragment-ion. The PIY of a fragment-ion $i$ was obtained as follows:

\begin{equation}
PIY_i = \left( \frac{A^{+}_i }{A^{+}_t}  \right) \times 100\% \pm \Delta PIY_{i}
\end{equation}

\begin{equation}
\Delta PIY_{i}= PIY_i \times  \sqrt{\left( \frac{\Delta A_{i}}{A^{+}_i}\right)^2  + \left( \frac{\Delta A_{t}}{A^{+}_t}\right)^2}
\end{equation}

where $A^{+}_i$ is the area of the gaussian profile fitted for the fragment-ion peak $i$, and $A^{+}_t$ is the total area of the PEPICO spectrum. The estimated determination, $\Delta PIY_i$, reflects the uncertainty in the area $\Delta A_{i}$ of each peak, and $\Delta A_{t}$ is the sum of these uncertainties. The uncertainties of the partial yields are estimated around 10\%. The PIYs for all the spectra are listed in Table \ref{tab:PIY}. For the monocations, only fragments with PIY greater than 1\% are shown. Based on the partial yields it can be concluded that the photodissociation of toluene leads preferentially to C$_{3}$H$_{3}^{+}$, C$_{3}$H$_{2}^{+}$, C$_{4}$H$_{2}^{+}$, C$_{4}$H$_{3}^{+}$, C$_{5}$H$_{3}^{+}$, C$_{2}$H$_{2}^+$, C$_{2}$H$_{3}^{+}$, and C$_{7}$H$_{7}^{+}$. For the doubly ionized C$_7$H$_n^{++}$ molecules, the total PIY contribution is of 1.8-2.6\%, between 277 eV and 302.7 eV. The most stable doubly charged ion is the parent molecule, C$_{7}$H$_{8}^{++}$, followed by C$_{7}$H$_{2}^{++}$ and C$_{7}$H$_{6}^{++}$, \textit{i.e.} closed-shell species with an even number of hydrogen atoms. The most produced dication with odd number of hydrogen is C$_{7}$H$_{7}^{++}$, followed by C$_{7}$H$_{3}^{++}$.

The absolute cross section determination is described elsewhere \cite{boe05}. Briefly, the non-dissociative single ionization (photoionization) cross section $\sigma_{ph-i}$ and the dissociative single ionization (photodissociation) cross section $\sigma_{ph-d}$ of toluene were determined by:

\begin{equation}
\sigma_{ph-i} = \sigma_{ph-abs} \frac{PIY_{C_{7}H_{8}^+}}{100}
\end{equation}
and
\begin{equation}
\sigma_{ph-d} = \sigma_{ph-abs} \Big( 1 - \frac{PIY_{C_{7}H_{8}^+}}{100} \Big)
\end{equation}
where $\sigma^{ph-abs}$ is the photoabsorption cross section \cite{ishii1987}.

The double photoionization cross-section of toluene, $\sigma_{ph-ii}$, and the cross-section of dications production, $\sigma_{++}$, were obtained as follows:

\begin{equation}
\sigma_{ph-ii} = \sigma_{ph-abs} \frac{PIY_{C_{7}H_{8}^{++}}}{100}
\end{equation}
and
\begin{equation}
\sigma_{++} = \sigma_{ph-abs} \frac{\sum\limits_{n=2}^{8} PIY_{C_{7}H_{n}^{++}}}{100}
\end{equation}

The photoabsorption, photoionization and photodissociation cross sections of toluene at the C1s edge are listed in Table \ref{tab-sigma}. The photoabsorption cross section of toluene at the C1s resonance energy is 1.9$\times 10^{-17}$ cm$^2$, almost 5 times higher than that of benzene. This higher value gives rise to higher photoionization and photodissociation cross sections for the molecule, and could be responsible for the higher production of doubly charged species of toluene, when compared to benzene.

\subsection{Theoretical Results}

\begin{figure}[t!]
\centering
  \includegraphics[width=70mm]{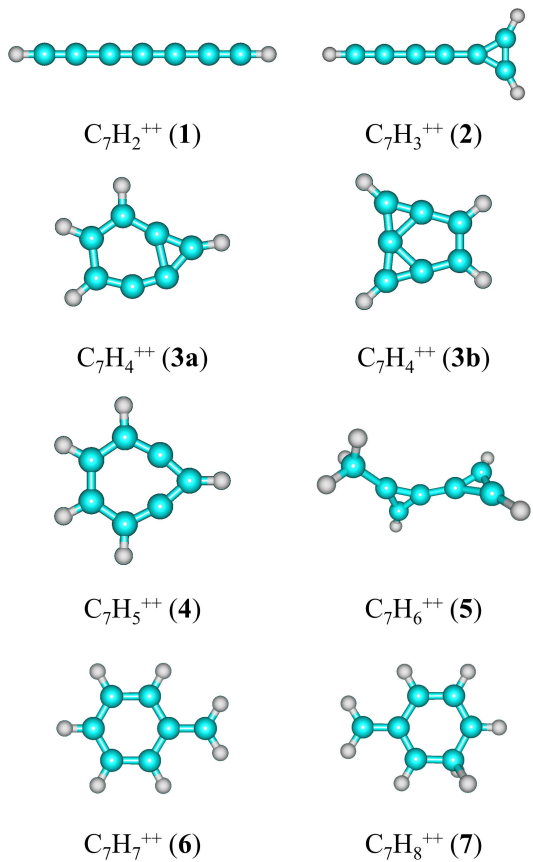}
\caption{Global minimum structures for the C$_7$H$_n^{++}$ species, n=2-8, at the M06-2X/cc-pVTZ(-f) level.} 
\label{fig:mol}
\end{figure}

Figure \ref{fig:mol} shows the most stable structures for each C$_7$H$_n^{++}$ species, with n varying from 2 to 8. It is worth mentioning that, although some of the dications were already studied from the theoretical point of view, to the best of our knowledge no comprehensive study concerning their structures and relative stabilities has been presented. The most stable C$_7$H$_2^{++}$ isomer is the all-linear heptatriynylidene dication (\textbf{1}), a doubly charged polyyne species. Although several neutral and anionic polyynes have been detected in interstellar clouds, such as C$_6$H and C$_6$H$_2$ \cite{ziu06}, C$_6$H$^-$ \cite{mcc06}, HC$_6$CN (Nguyen-Q-Rieu et al. 1984), C$_8$H (Cernicharo \& Gu\'{e}lin 1996), C$_8$H$^-$ (Br\"{u}nken et al. 2007), and HC$_{10}$CN \cite{bel97}, no dicationic analogue has been detected so far. Polyynes were also suggested to be present in interstellar sources attached as substituents on PAHs \cite{dul09}. The most stable C$_7$H$_3^{++}$ isomer, on the other hand, is the butadiynylcyclopropenyl dication (\textbf{2}), a planar hydrocarbon species containing a linear polyyne attached to a three-membered ring. 

For C$_7$H$_4^{++}$, a structure containing a planar tetracoordinate carbon (ptC, \textbf{3b}) is only 0.5 kcal mol$^{-1}$ higher in energy than the cyclic \textbf{3a} structure. At this level of calculation, it is not possible to discriminate which of the two structures is the ground state for the C$_7$H$_4^{++}$ system. This problem will be addressed in a future study dealing with the low-lying isomers of the C$_7$H$_n^{++}$ dications. 

For C$_7$H$_6^{++}$, the ground state structure is usually attributed in the literature to that of the cycloheptatrienylidene dication \citep{roi09}. However, a previously non-reported methylbiscyclopropenyl structure (\textbf{5}) was predicted in this work as the global minimum for the system. Its energy is 6.5 kcal mol$^{-1}$ smaller than the usual seven-membered ring. Finally, for the the parent dication C$_7$H$_8^{++}$, the meta-protonated benzyl cation (\textbf{7}) is the most stable structure, highlighting the loss of structural integrity from neutral toluene, in agreement with previous works \cite{roi06}.

\begin{table}[t]
\centering
\caption{Electronic energy corrected by ZPE ($\Delta$E$_0$) of selected dissociation pathways, compared to the estimated coincidence from PE2PICO spectra. Energy values are in kcal mol$^{-1}$; yields are in \%.}
\label{tab:zpe}
\begin{small}
\begin{tabular}{llll}
\hline
\hline
\rule{0pt}{3ex}Dication        & Dissociation Pathway        & $\Delta$E$_0$ (Yield)                  &  \\
\rule{0pt}{3ex}C$_7$H$_5^{++}$ & C$_3$H$_3^+$ + C$_4$H$_2^+$ & -34.9 (4.06)                           &  \\
\rule{0pt}{3ex}C$_7$H$_7^{++}$ & C$_3$H$_3^+$ + C$_4$H$_4^+$ & -31.1 (2.09)                           &  \\
\rule{0pt}{3ex}C$_7$H$_6^{++}$ & C$_3$H$_3^+$ + C$_4$H$_3^+$ & -23.3 (3.91)                           &  \\
\rule{0pt}{3ex}C$_7$H$_5^{++}$ & C$_2$H$_2^+$ + C$_5$H$_3^+$ & -11.4 (3.16)                           &  \\
\rule{0pt}{3ex}C$_7$H$_8^{++}$ & C$_2$H$_3^+$ + C$_5$H$_5^+$ & -4.6 (1.86)                            &  \\
\rule{0pt}{3ex}C$_7$H$_5^{++}$ & C$_3$H$_2^+$ + C$_4$H$_3^+$ & -3.8 (3.56)                            &  \\
\rule{0pt}{3ex}C$_7$H$_6^{++}$ & C$_2$H$_3^+$ + C$_5$H$_3^+$ & -3.7 (4.47)                            &  \\
\rule{0pt}{3ex}C$_7$H$_8^{++}$ & CH$_3^+$ + C$_6$H$_5^+$     & 2.7 (1.05)                             &  \\
\rule{0pt}{3ex}C$_7$H$_5^{++}$ & C$_2$H$_3^+$ + C$_5$H$_2^+$ & 8.6 (1.41)                             &  \\
\rule{0pt}{3ex}C$_7$H$_7^{++}$ & C$_2$H$_3^+$ + C$_5$H$_4^+$ & 18.1 (1.00)                            &  \\
\rule{0pt}{3ex}C$_7$H$_7^{++}$ & CH$_3^+$ + C$_6$H$_4^+$     & 24.6 (1.05)                            &  \\
\rule{0pt}{3ex}C$_7$H$_4^{++}$ & C$_3$H$_2^+$ + C$_4$H$_2^+$ & 25.2 (4.88)                            &  \\
\rule{0pt}{3ex}C$_7$H$_4^{++}$ & C$_2$H$_2^+$ + C$_5$H$_2^+$ & 41.5 (2.67)                            &  \\
\rule{0pt}{3ex}C$_7$H$_3^{++}$ & C$_3$H$^+$ + C$_4$H$_2^+$   & 45.2 (2.71)                            &  \\
\rule{0pt}{3ex}C$_7$H$_3^{++}$ & C$_2$H$_2^+$ + C$_5$H$^+$   & 48.9 (2.00)                            &  \\
\rule{0pt}{3ex}C$_7$H$_3^{++}$ & C$_3$H$_2^+$ + C$_4$H$^+$   & 68.2 (1.45)                            &  \\
\rule{0pt}{3ex}C$_7$H$_2^{++}$ & C$_3$H$^+$ + C$_4$H$^+$     & 93.5 (1.73)                            & \\
\hline
\hline
\end{tabular}
\end{small}
\end{table}

The thermodynamics of selected dissociation pathways by Coulombic decay from C$_7$H$_n^{++}$ dications was also studied. The dissociation pathways were chosen from the PE2PICO analysis, and their experimental yields were compared to the variation of electronic energy corrected by zero point energy (ZPE). From this analysis, the viability of the pathways proposed by the experiments  was evaluated.

\begin{figure*}[t!]

\centering
  \includegraphics[width=120mm]{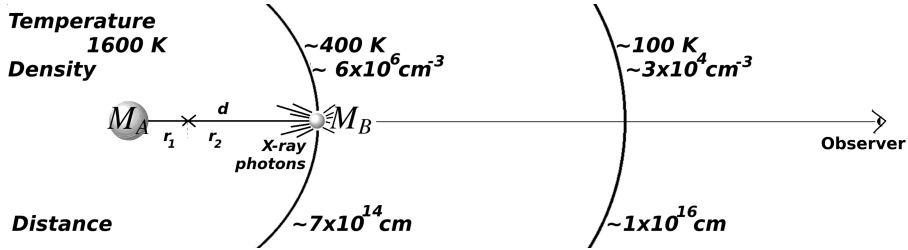}
\caption{Sketch of the geometry adopted. M$_{A}$ is the AGB star and M$_{B}$ your companion. The X-ray flux towards the AGB star is given by r$_{2}$ and the distance from the AGB star is given by r$_{1}$. The distance between the stars is given by d = r$_{1}$ + r$_{2} \sim$ 45 AU. This distance is the result of a X-ray flux passing through a column density of 4$\times$10$^{21}$ cm$^{-2}$ in the direction to the observer.} \label{fig:geometry}
\end{figure*}

Table \ref{tab:zpe} shows the ZPE corrected electronic energy ($\Delta$E$_0$) of selected dissociation pathways in comparison with the estimated coincidence from the PE2PICO spectra. For more hydrogenated C$_7$H$_n^{++}$ dications (n=5-8), there is a reasonable correlation between the lower energy pathways and the coincidence of monocations from the PE2PICO analysis. The loss of correlation for less hydrogenated C$_7$H$_n^{++}$ dications (n=2-4) suggests that part of the estimated yield from the PE2PICO spectra could be associated with pathways in which there is also the formation of neutral species. Nevertheless, the most exoergic dissociation pathways are associated to the loss cyclic C$_3$H$_3^+$, which also appears as the most prominent peak in the PEPICO spectra, although the presence of the less stable propargyl isomer cannot be ruled out as it could be formed in the experiment and, once formed, the large activation barrier could prevent its isomerization to the cyclic structure \cite{ric10}. This result suggests that a significant part of the formation of C$_3$H$_3^+$ comes from the decomposition of C$_7$H$_n^{++}$ dications, previously generated by Auger mechanism. The cyclopropenyl ion, C$_3$H$_3^+$, is considered the simplest H\"{u}ckel aromatic system, having 2$\pi$ electrons. It is suggested that C$_3$H$_3^+$ plays an important role in the atmospheric chemistry of Titan \cite{ali13}. Since toluene is also found in the upper atmosphere of Titan, part of the production of C$_3$H$_3^+$ must come from the photodissociation of toluene.

\subsection{Ions in the Circumstellar Envelope of AGB Stars}

As mentioned before, the production of large PAHs is mainly attributed to the chemical process occurring in the circumstellar envelope of AGB stars. T Dra, a carbon-rich AGB star, is at a distance of 610 pc, with bolometric luminosity L = 6300 L$_{\odot}$, effective temperature T$_{*}$= 1600 K, velocity of expanding gas v$_{exp}$ = 13.5 km s$^{-1}$, and mass loss \.{M} = 1.2 $\times$ 10$^{-6}$ M$_{\odot}$/yr \cite{Schoier2001}. The soft X-ray spectrum (0.2-2 keV) of the object shows emission centered at about $\lesssim$ 1 keV, observed by ROSAT \cite{Ramstedt2012}. In order to obtain the photon flux as a function of the energy, a value of L$_{X}$ $\approx$ 3.2$\times$10$^{31}$ erg s$^{-1}$ was used for the X-ray luminosity, corresponding to a column density of N$_{H}$ $\approx$ 4,0$\times$10$^{21}$ cm$^{-2}$ \cite{Ramstedt2012}.

The main suggestions of Ramstedt et al. (2012) for the origin of the X-ray emission of T Dra are that it is due: 1) to the mass-accretion onto a compact white-dwarf, in the case of binarity, or 2) to a coronal emission from a large scale magnetic field. The gas density nearby the stellar surface should be sufficient to absorb the X-ray produced by the surface magnetic field, but morphological asymmetries could affect the photoabsorption and allows X-ray emission. Here, we are considering the case of binarity. In fact, it has been suggested that the morphological structure of CRL 618 and IRC +10216, two well studied evolved stars, are due to binary companions \cite{Cernicharo2015, Velazquez2014}. The numerical simulations of Val-Borro (2009) of mass-accretion of wind onto an AGB companion suggest this scenario as reasonable. The position of the companion in the extended expanding envelope of the AGB star, which could result in the X-ray absorption constrained by Ramstedt et al. 2012, does not require any more special morphological structure, and it is used as the scenario in this work (Figure \ref{fig:geometry}).

Following the equations of Glassgold (1996) for a spherical symmetry and a uniform gas expansion, with column density $\alpha$ r$^{-1}$, the distance between the AGB star and the small companion that results in a column density of 4.0$\times$10$^{21}$ cm$^{-2}$ observed from Earth should be $\sim$ 45 AU. 

The X-ray flux, $F_{X}(E)$, attenuated while it penetrates the gas envelope, is given by:

\begin{equation}
 F_{X}(E) = \frac{L_{X}}{4\pi r_{2}^{2} h\nu}e^{-\tau}
\end{equation}

where $r_{2}$ is the distance from the companion, and $\tau$ is the optical depth:

\begin{equation}
\tau = \sigma_{X}N_{H} 
\end{equation}

where $\sigma_{X}$ is the total X-ray photoabsorption cross-section by gas in the interstellar medium, and N$_H$ is the hydrogen column density. Considering the photon flux in the vicinity of the C1s resonance energy obtained from Draine $\&$ Tan (2003), we evaluate that $\sigma_{X}$ is $\sim$ 2.3$\times$10$^{-21}$ cm$^{2}$. The N$_H$ value at a given distance from the AGB's companion was obtained using the equation (2) of Glassgold (1996):

\begin{equation}
N_{H}(r_{1}) = 2,67\times10^{36} \left[ \frac{1}{d}- \frac{1}{r_{1}}\right] \texttt{cm}^{-2},
\end{equation}

where d is the distance between the stars and r$_{1}$ is the distance from the AGB star.

\begin{figure}[t!]
\centering
  \includegraphics[width=75mm]{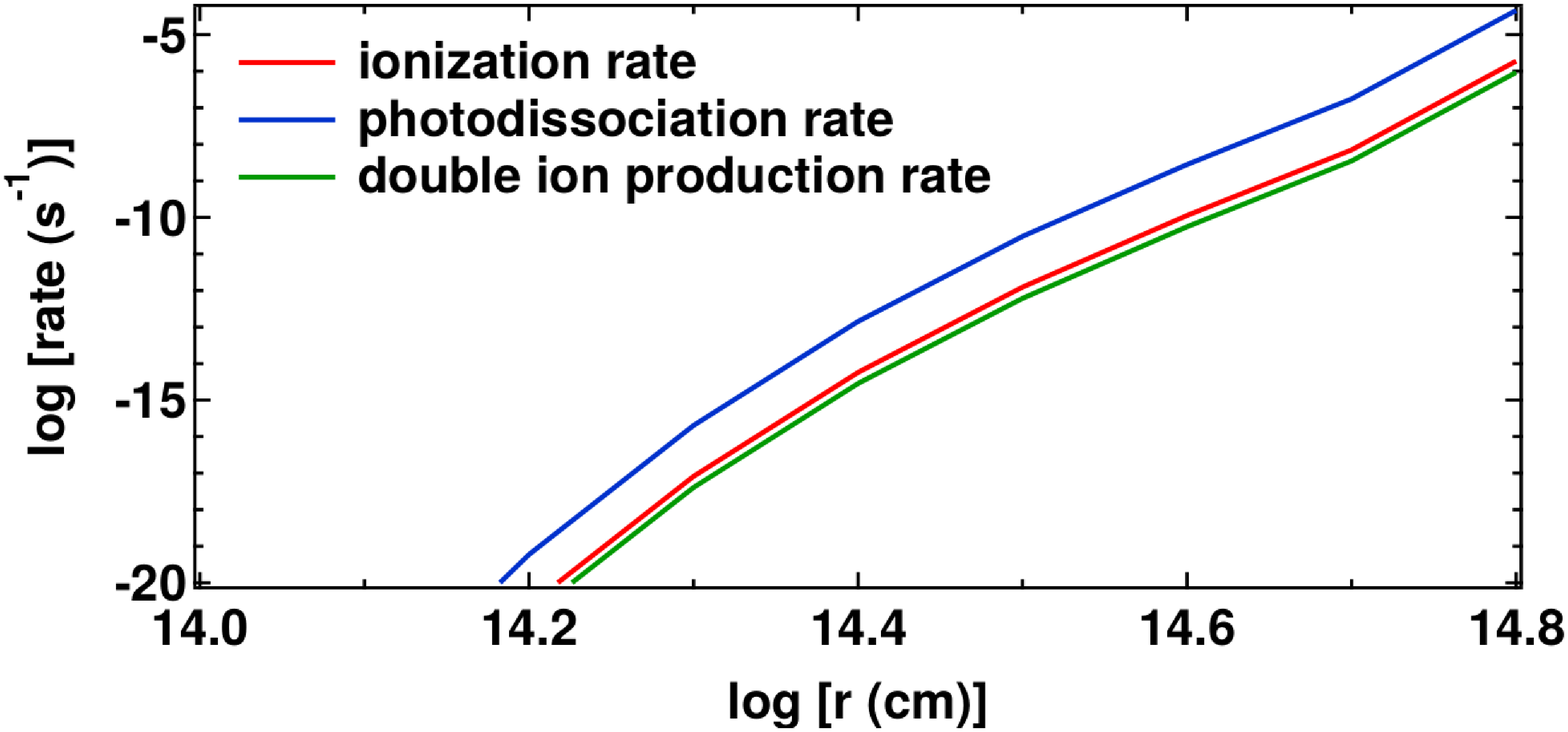}
\caption{Photoionization and photodissociation rates of toluene and photoproduction of dications as functions of the distance from the AGB star.} \label{fig:iondiss}
\end{figure}

With these parameters, we can evaluate the rates of photodissociation, ionization and half-life of toluene, besides the production of double charged ions through the T Dra envelope that faces the companion star. The dissociation of a given molecule subjected to a radiation field is given by:

\begin{equation}
-\frac{dN}{dt}=Nk_{ph-d},
\label{phdcs}
\end{equation}

where the photodissociation rate, $k_{ph-d}$, is defined as:

\begin{equation}
 k_{ph-d} = \sigma_{ph-d}(E)F_{X}(E).
 \label{phdcs}
\end{equation}

On the other hand, the photoionization rate is given by:
\begin{equation}
\varsigma_{i}= \sigma_{i}(E)F_{X}(E).
\end{equation}

and the rate of dications production is:
\begin{equation}
\varsigma_{i}= \sigma_{++}(E)F_{X}(E).
\end{equation}

The results are shown in the Figure \ref{fig:iondiss}. The photoionization, photodissociation and photoproduction of dications drops sharply in the direction of the inner envelope. A half-life radial profile of toluene can be evaluated from the equation \ref{phdcs}, by writing $t_{1/2} = ln2/k_{ph-d}$, that gives 10$^{6}$ years of half-life at a distance of 3$\times$10$^{14}$ cm (21 AU) from the AGB star. Further chemical reactions should be inhibited as the shell expands outwards, creating a photodissociation radius determined by the molecular half-life. The production of molecules is, thus, strongly dependent on the side of the envelope that is facing the companion star in a binary case. Nevertheless, if magnetic activity is present through the envelope, the ions produced by these photochemical reactions may contribute to ambipolar diffusion. Therefore, the structural diversity of the single and double ions would be shared through the envelope within a new network of ion-molecule reactions in the expanding shells of the entire envelope of the AGB star, which has not yet been explored by any chemical models.

\section{Summary and conclusions}

In this work, the X-ray photoionization and photodissociation of toluene was studied in the vicinity of the carbon K shell resonance energy. It was shown that the main photodissociation pathways lead to the formation of C$_{3}$H$_{3}^{+}$, C$_{3}$H$_{2}^{+}$, C$_{4}$H$_{2}^{+}$, and C$_{4}$H$_{3}^{+}$. Moreover, electron-ion coincidences have revealed several doubly charged molecular ions containing seven carbon atoms with considerable abundance. The relatively high survival of the dications subjected to hard inner shell ionization suggests that they could be observed in the interstellar medium, especially in regions where PAHs are detected, and validates their search in astrophysical environments.

The absolute single and double photoionization and photodissociation cross sections were determined. The particularly high photoabsorption cross section of toluene at the C1s edge increases the photoionization and photodissociation cross sections of the molecule, when compared to benzene. Eventually, this could be responsible for the higher detection of doubly charged species of toluene.

The structure of the C$_7$H$_n^{++}$ dications was also studied in this work. It was shown that, while C$_7$H$_2^{++}$ and C$_7$H$_3^{++}$ present polyyne-like structures, a cyclic structure containing a planar tetracoordinate carbon is one of the possible global minima of C$_7$H$_4^{++}$. A previously non-reported methyl-biscyclopropenyl structure was found as the global minimum for C$_7$H$_6^{++}$, and the loss of structural integrity of C$_7$H$_8^{++}$ when compared to neutral toluene, as previously suggested in the literature, was confirmed.

The most exoergic dissociation pathways from doubly charged ions are associated with the loss of c-C$_3$H$_3^+$. This suggests that the decomposition C$_7$H$_n^{++}$ dications could contribute to the enhancement of the cyclopropenyl ion in toluene-rich astrophysical environments, such as the upper atmosphere of Titan. 

Finally, the photoionization and photodissociation cross sections obtained in this work were applied to evaluate the half-life of toluene through the T Dra envelope, which reaches a value of $\sim$10$^{6}$ years at a distance of 21 AU from the AGB star. The results also suggest that the ion production on the side of the envelope that faces the companion white dwarf could affect other regions, if magnetic activity is present. Ultimately, this mechanism could trigger ion-molecule reactions through ambipolar diffusion.

\acknowledgments

The authors would like to thank CNPq, CAPES, and FAPERJ for financial support.


{\it Facilities:} \facility{Laborat\'{o}rio Nacional de Luz S\'{i}ncrotron (LNLS)}, \facility{Laborat\'{o}rio de Qu\'{i}mica Te\'{o}rica e Modelagem Molecular (LQTMM/IQ-UFRJ)}.

\clearpage

\end{document}